\documentclass[12pt]{article}
\usepackage{psfig}
\usepackage{graphicx}
\usepackage{citesupernumber}
\makeatletter

\def\frtnsfb{\fontfamily{\sfdefault}\fontseries{bx}\fontshape{n}\fontsize{14.4}{18pt}\selectfont}
\def\svtnsfb{\fontfamily{\sfdefault}\fontseries{bx}\fontshape{n}\fontsize{17.28}{22pt}\selectfont}
\def\twtysfb{\fontfamily{\sfdefault}\fontseries{bx}\fontshape{n}\fontsize{20.74}{25pt}\selectfont}
\def\twfvsfb{\fontfamily{\sfdefault}\fontseries{bx}\fontshape{n}\fontsize{24.88}{30pt}\selectfont}
\def\large{\@setsize\large{18pt}\xivpt\@xivpt \let\sfb=\frtnsfb}
\def\Large{\@setsize\Large{22pt}\xviipt\@xviipt \let\sfb=\svtnsfb}
\def\LARGE{\@setsize\LARGE{25pt}\xxpt\@xxpt \let\sfb=\twtysfb}
\def\huge{\@setsize\huge{30pt}\xxvpt\@xxvpt \let\sfb=\twfvsfb}
\makeatother
\begin{document}
\parindent=10mm
\baselineskip=8mm

\bibliographystyle{nature}
\noindent {\Large\sfb Path Coalescence in Spatially Correlated Random\\
Walks}\\[4.5mm]
\noindent { Michael Wilkinson$^{\ast}$ and Bernhard Mehlig$^{\ast\ast}$}\\[4.5mm]
\noindent {\em \mbox{}$^{\ast}$Faculty of Mathematics and Computing, 
The Open University, Walton Hall, Milton Keynes, MK7 6AA, England,
\mbox{}$^{\ast\ast}$School of Physics and Engineering Physics, Gothenburg University/Chalmers,
Gothenburg, Sweden.}\\[4.5mm]

\noindent {\bf 
A particle subject to  successive, random displacements is
said to execute a random walk (in position or some other coordinate).
The mathematical properties of random walks 
have been very thoroughly investigated, and the model is used in 
many areas of science and engineering\cite{vKa81,Sch76} as well as 
other fields such as finance\cite{Wil95} and the life sciences\cite{Kimura}.
This letter describes a phenomenon occurring in a natural extension 
of this model: we consider the motion of a large number of particles 
subject to successive random displacements which are correlated in space, 
but not in time. If these random displacements are smaller than their 
correlation length, the trajectories coalesce onto a decreasing number 
of trails. This surprising effect is explained and quantitative results
are obtained. Various possible realisations are discussed, ranging from 
coalescence of the tracks of water droplets blown off a windshield to 
migration patterns of animals.}\\[4.5mm]

The phenomenon we have discovered is illustrated in figure~\ref{fig1}. It 
shows the positions $x$ of $20$ particles as a function
of time $t$. Each particle undergoes a random walk, but
the displacements of nearby particles are correlated,
such that particles which are very close experience almost
identical displacements. The paths of the particles
are seen to coalesce onto a decreasing number of trails,
along which the particles follow almost the same trajectory.
This effect is surprising because random walks normally
reduce rather than accentuate inhomogeneities in the density 
of particles. 
In the following we explain this effect quantitatively
and suggest some contexts in the physical
sciences in which the effect can certainly be observed, and 
areas in biology where it may find interesting 
applications. 

We consider a system of particles labelled by an index $i$, 
having positions $x_i(t)$ at time $t$. 
The particles are subjected to random displacements
at times which are integer multiples of a small increment
$\delta t$. At time $t=n\,\delta t$ the displacement is given by a 
random function $f_n(x)$ evaluated at the position $x_i(t)$:
\begin{equation}
\label{eq1}
x_i(t\!+\!\delta t)=x_i(t)+f_n\big(x_i(t)\big)
\ .
\end{equation}
The random displacements $f_n$ satisfy
$\langle f_n(x)\rangle=0$
and $ \langle f_n(x)f_{n'}(x')\rangle=\delta_{nn'}c(x-x')$.
Here $\delta_{nn'}$ is 
unity if $n=n'$
and zero otherwise, and we write
$\langle A \rangle$ for the ensemble 
average of a quantity $A$ (that is, the average over different
trials of the random process generating $A$).
The correlation function $c(X)$ is even in $X=x-x'$, 
decays rapidly as $X\to \infty$,
and is small when $\vert X\vert \gg \xi$. 
A suitable choice (adopted in figure \ref{fig1} and in the following)
is $c(X)=\varepsilon^2\exp(-X^2/2\xi^2)$,
$\varepsilon$ is the typical magnitude of each displacement and
$\xi$ is the correlation length.
Figure \ref{fig1} used the values
$\varepsilon^2\approx8\times 10^{-6}$ and  $\xi= 10^{-1}$.

As an example of a situation where the effect could be observed,
consider the motion of liquid droplets on a surface, 
moving in one direction under
a constant force (rain blown off a perspex windshield is an example
of this situation). If the surface is randomly contaminated, the wetting
angle will be different on opposite sides of each drop, and the trajectory
of the droplet will be randomly deflected. We are concerned with
the case where the surface contaminants are smeared over an area
large compared to the droplets 
(perhaps resulting from cleaning the windshield with a waxy polish), 
so that nearby droplets are deflected
in the same direction. There is no interaction between the drops unless
they are close enough to combine due to surface tension: we stress
that the coalescence is that of the paths taken by different drops,
not of the droplets themselves. We model the motion of a droplet
across the surface by a particle of mass $m$.
At position ${\bf r}=(x,y)$ on the surface, the drop is 
subject to a force ${\bf F}({\bf r})+F_0\, {\bf j}$, where $F_0$ is the 
magnitude of a steady force acting in the direction of the unit 
vector ${\bf j}$ defining the $y$-axis and ${\bf F}({\bf r})$ 
is a homogeneous and isotropic 
random force with correlation length $\xi$. 
We assume that the particles are subjected to a viscous 
resistive force proportional to their velocity across the surface, 
such that the equations of motion are
\begin{equation}
\label{eq2}
m{{\rm d}{\bf r}\over{{\rm d}t}}={\bf p}\ ,\ \ \ 
{{\rm d}{\bf p}\over{{\rm d}t}}
=F_0 \,{\bf j}+{\bf F}({\bf r})-\gamma {\bf p}
\end{equation}
(where ${\bf p}$ is the momentum of the drop).
When the fluctuating force is weak, the 
trajectories are locally approximated by straight lines, 
with $x$ approximately constant and 
with $y$ increasing at a rate $v_y=F_0/m\gamma$. Furthermore,
when the damping $\gamma $ is large\cite{note1}, the velocity is given 
by the local value of the force: ${\bf v}(t)\sim {\bf p}(t)/m=v_y{\bf j}
-{\bf F}(x(t),v_yt)/m\gamma$. 
Under these circumstances 
the displacement $x(t)$ perpendicular to the steady force satisfies
\begin{equation}
\label{eq3}
{{\rm d}x(t)\over{{\rm d}t}}
={1\over {m\gamma}}
F_x\bigl(x(t),v_yt\bigr)
\ .
\end{equation}
When $v_y$ is large, the velocity of the particle fluctuates very 
rapidly, and in this limit it is reasonable to model equation (\ref{eq3}) 
by the stochastic equation (\ref{eq1}).

Why do the trajectories in figure~\ref{fig1} merge?
The simplest argument is based upon linear stability
analysis. The separation $\delta x(t)$ of two nearby 
trajectories 
varies exponentially in time,
with rate
$\lambda = t^{-1}\,\langle\log | \delta x(t) |/|\delta
x(0)|\rangle
={\langle \log \vert 1+f'_n\vert \rangle /{\delta t}}\,.$
Provided $\varepsilon/\xi$ is small, the magnitudes of the derivatives
$f'_n ={\rm d}f_n/{\rm d}x$ are small compared to unity. Taylor expansion 
of the logarithm gives 
$\lambda \sim -{1\over 2}\langle {f_n^\prime}^2\rangle/\delta t$, so that 
 $\lambda $ is negative. 
In this case, most nearby trajectories approach each other with an exponentially
decreasing separation, implying coalescence.
This argument also indicates 
that if $\varepsilon/\xi$ is large, the coalescence effect disappears: 
if the function $f_n$ has a Gaussian 
distribution, the exponent $\lambda $ becomes positive when 
$\langle {f_n^\prime}^2\rangle$ exceeds $2.421...$.

A more complete and concrete understanding of the coalescence 
effect is obtained by considering statistics of the density
of particles, $\varrho(x,t)=\sum_i \delta \bigl(x_i(t)-x\bigr)$.
Translational invariance implies that an initially uniform 
density
remains uniform,  $\langle \varrho(x,t)\rangle=\varrho_0$.
Path coalescence 
is revealed 
by the density-density correlation function 
${\cal K}(x,x';t) = \big\langle \varrho(x,t)\,\varrho(x',t)\big\rangle -
\varrho_0\delta(x-x')$.
Because of translational invariance,
the correlation function ${\cal K}$ is a function of $X=x-x'$ only: 
we write ${\cal K}(x,x';t)=K(x-x',t)$. 
A tendency for particles to cluster is demonstrated
by ${K}(X,t)$ becoming large for $X$ small, in the limit $t\to \infty$.
When the typical magnitude
of the jumps is small compared to the correlation
length $\xi$, we find that the correlation function satisfies a generalised
diffusion equation, or Fokker-Planck equation\cite{vKa81,Sch76}
\begin{equation}
\label{eq4}
{\partial K(X,t)\over{\partial t}}={\partial^2\over{\partial X^2}}
\bigl[D(X)\,K(X,t)\bigr]\,.
\end{equation}
The diffusion constant 
$D(X)=[c(0)-c(X)]/\delta t$
approaches zero quadratically
at the origin: $D(X)\sim \kappa X^2$ for $X\ll \xi$,
where $\kappa=-{1\over 2}c''(0)/\delta t$. When $X \gg \xi$
the diffusion constant approaches a constant value, 
$D_0=\varepsilon^2/\delta t$. Equation (\ref{eq4}) was obtained from 
the stochastic model, equation (\ref{eq1}), but in the 
appropriate limits the density-density correlation 
function of (\ref{eq3}) also satisfies\cite{note2}
equation (\ref{eq4}).

Now consider the properties of solutions of 
equation (\ref{eq4}). We note that equation (\ref{eq4})
is in the form of a continuity equation, 
$\partial K/\partial t+\partial J/\partial x=0$, so that the
integral of the correlation function over all $X$ is a conserved
quantity. The flux of the correlation function passing the
separation parameter $X$ at time $t$ is
\begin{equation}
\label{eq5}
J(X,t)=-{\partial \over{\partial X}}\big[D(X)K(X,t)\big]
\ .
\end{equation}
Consider an initially uniform distribution of density, with
value $\varrho_0$
(corresponding to $K(X,0)=\varrho_0^2$). For $X\ll \xi$ 
the diffusion constant is an increasing function of $X$. 
Together with 
(\ref{eq5}) 
this implies an initial flux of 
correlation towards $X=0$. At large times, 
$K(X,t)$ is thus sharply peaked at the origin. 
For $X\gg \xi$, on the other hand, the approximate solution of 
(\ref{eq4}) is
\begin{equation}
\label{eq6}
K(X,t)\sim \varrho_0^2\,
{\rm erf}\biggl({\vert X\vert \over{\sqrt{4D_0 t}}}\biggr)
\ .
\end{equation}
Using the fact that $K(X,t)$ satisfies a conservation
law, we deduce that the average number of particles condensing
into each trail at time $t$ is
\begin{equation}
\label{eq7}
N(t)\sim {4\over{\sqrt{\pi}}}\varrho_0\sqrt{D_0 t}
\ .
\end{equation}
When $t/\delta t$ is large, and
$X \ll \xi$ (but not too close to zero)
the flux $J(X,t)$ is found to be approximately uniform. This implies
\begin{equation}
\label{eq8}
K(X,t)\sim {\varrho_0^2\over{\kappa}}\sqrt{D_0\over{\pi}}{1\over{X\sqrt t}}
\ .
\end{equation}
The $1/X$ divergence of (\ref{eq8})
is non-integrable, so that this expression must fail 
near the origin.
An exact calculation shows that $K(X,t)$ is finite at $X=0$, 
the height of the peak is 
$K(0,t)=\exp(2\kappa t)$. 
In summary, at large times the correlation function $K(X,t)$ develops
a correlation hole for $\xi \ll X \ll \sqrt{4 D_0 t}$
reflecting the path coalescence effect.
The results of numerical simulations shown in figure~\ref{fig2}
confirm these predictions.

In the remainder we discuss 
a number of possible 
realisations of the path coalescence 
effect. Unlike the 
case of liquid droplets described above, these are
speculative, but they show that the path coalescence effect
is likely to have a very broad range of applications. 

There are many potential applications in the physical sciences
involving fast particles interacting with a random potential.
Two examples of this type are
the motion of rocks rolling down a scree slope,
and the motion of highly energetic electrons 
through disordered crystalline solids. 
There are also possible connections between the coalescence
effect and a \lq streaming'
observed in the flow of electrons away from a constriction
in a two-dimensional electron gas with very low scattering\cite{Top01},
and we shall discuss this case in some detail.
The experiment shows regions of markedly increased current
density persisting to some distance from the constriction.
This was explained by showing similarity
to simulations of independent electron 
motion in the smoothly varying random potential
of the doping atoms. This model is essentially (\ref{eq2}), 
with $\gamma=0$. 
Theoretical discussion of this system\cite{Kap02} has emphasised that 
caustics are important in understanding the 
empirical results. 
We remark that these experiments might show an even more 
pronounced effect if dissipation
were introduced. Equations (\ref{eq2}) can show various
types of behaviour; figure~\ref{fig3} shows numerically 
computed trajectories for one choice of parameters. 
These trajectories initially show evidence of the streaming effect 
discussed in references\cite{Top01,Kap02} with 
the associated caustics. For larger times the trajectories coalesce, 
in contrast to the dissipationless
model where the streams eventually disperse. We note that 
dissipation of the electron motion could be increased by
increasing the temperature of the system. 

There are also potential applications in the biological
sciences, involving the movement of organisms in response
to small random fluctuations in their environment. Our
model provides a mechanism through which large
numbers of organisms can congregate without communicating.

One example of this type is the migration
of animals across a nearly homogeneous smooth terrain. 
Thus figure~\ref{fig1} could be thought of as a map showing paths
of animals on an Eastward migration.
The paths of the animals will be deflected by small
random fluctuations of topography or vegetation. 
Our calculations show that the animals can be drawn 
together onto the same paths,
even if there is no communication between them, and
no gross features in the terrain favouring particular
routes. 

A second example applies to simple   
organisms such as plankton which can move in response to
changes in their environment, such as nutrient concentration. 
In cases where there
are small, spatially correlated random fluctuations of the 
nutrient concentration, the path coalescence effect could lead
to unexpectedly large concentrations of organisms. Such
a mechanism could be utilised by evolution, enabling simple organisms which 
cannot communicate directly to congregate for sexual reproduction.

\noindent {\bf Acknowledgement}  Anders Eriksson
(Gothenburg University) 
suggested to us that the path coalescence effect might be 
relevant to animal migration patterns.\\[4.5mm]
{\bf Correspondence} should be addressd to M. W.  
(e-mail: m.wilkinson@open.ac.uk) or
to B. M. (e-mail: mehlig@fy.chalmers.se).

\newpage

\begin{figure}
\caption{\label{fig1}
Positions $x_i(t)$ for $20$ particles performing spatially correlated 
random walks. The trajectories coalesce.}
\end{figure}

\begin{figure}
\caption{\label{fig2}
Statistics of the particle density $\varrho(x,t)$.
{\bf a}
The density-density correlation function $K(X,t)$ of the motion
illustrated in figure~\ref{fig1}, derived from simulations 
$(\circ)$, 
compared
with the limiting theoretical forms (\ref{eq6}), full line, 
and (\ref{eq8}), dashed line. 
Parameter values:
$\varepsilon^2 
\approx 1.25\times10^{-8}$, $\xi \approx 6.4\times10^{-3}$, 
and $t=5\times 10^5\delta t$. 
The inset shows numerical results verifying
that $K(0,t)=\exp(2\kappa t)$ for the same parameter values.
 {\bf b}
The mean number of particles in a cluster, $N(t)$, for the process
shown in figure \ref{fig1}. The results of the simulation ($\circ$) 
are compared with theory (\ref{eq7}), full line.
Parameter values:
$\varepsilon^2 \approx 2\times10^{-7}$, 
$\xi \approx 1.6\times10^{-3}$. Particles are considered
to be part of a cluster of $N$ if all of their positions
are within an interval of length $\xi$.
}
\end{figure}

\begin{figure}
\caption{\label{fig3}
Trajectories (red) for damped motion in a smooth random potential 
described by equation (\ref{eq2}). 
Particles are introduced with 
uniform density on the left; here the $x$ axis is vertical.
The force is a sum of two terms: 
a random force derived from the gradient of a potential, 
${\bf F}({\bf r})=-\nabla V({\bf r})$ with 
$\langle V({\bf r})V({\bf r}')\rangle
=\varepsilon^2\exp[({\bf r}-{\bf r}')^2/2\xi^2]$,
and a steady force $F_0\, {\bf j}$ acting in the $y$-direction.
The potential is shown in green, higher values correspond
to darker colours.
Parameter values: $F_0=1.5$, $\varepsilon^2 =1.25\times10^{-4}$, 
$\xi\approx 6\times 10^{-2}$, $m=1$, and $\gamma=4$.  }
\end{figure}
\vfill
\flushbottom

\begin{thebibliography}{5}
\bibitem{note1}
 The condition defining this 
 {\sl overdamped} situation is $\gamma \tau\gg 1$, where $\tau$
 is the timescale of fluctuations of the potential experienced
 by a particle moving with speed $v_y$, namely $\tau=\xi/v_y$.
\bibitem{note2}
Equation (\ref{eq1}) is the It$\hat {\rm o}$ stochasticisation of equation
(\ref{eq3}).
If we use the arguably more appropriate Stratonovich
stochasticisation, we still arrive at the same equation (\ref{eq6})
for the correlation function.
\end{thebibliography}
\end{document}